\documentclass[11pt]{article}
\usepackage{amssymb}
\usepackage{amsfonts}
\usepackage{graphics}
\begin{document}
\begin{titlepage}

October 2004 \hfill{UTAS-PHYS-2004-05}\\
%\mbox{}\hfill{physics/04xxxxx}
\vskip 1.6in
\begin{center}
{\Large {\bf  A base pairing model of duplex formation I: Watson-Crick pairing geometries}}
\\[5pt]
\end{center}

\normalsize
\vskip .4in

\begin{center}
J.D. Bashford and P.D. Jarvis\\
{\it School of Mathematics and Physics, University of Tasmania} \\
{\it PO Box 252-37, Hobart 7001, Tasmania Australia} \\
%\par \vskip .1in \noindent
%

\end{center}
\par \vskip .3in

\begin{center}
{\Large {\bf Abstract}}\\
\end{center}
We present a base-pairing model of oligonuleotide duplex formation and 
show in detail its equivalence to the Nearest-Neighbour dimer methods from
fits to free energy of duplex formation data for short DNA-DNA and DNA-RNA 
hybrids containing only Watson Crick pairs. 
In this approach the connection between rank-deficient polymer  
and rank-determinant oligonucleotide parameter, sets for DNA duplexes
is transparent. The method is generalised to include RNA/DNA hybrids
where the rank-deficient model with 11 dimer parameters in fact provides 
marginally improved predictions relative to the standard method with 16 
independent dimer parameters ($\Delta G$ mean errors of 4.5 and 5.4 \% respectively).

%\end{center}
 
\vspace{3cm}
\end{titlepage}

\section{Introduction}
Simple nearest-neighbour (NN) models of helix stability at the base dimer
level have been refined over some years \cite{t2}-\cite{hyb2} and are 
commonly used, for example, to predict RNA secondary structure formation 
\cite{sl2}, \cite{mathews} and duplex melting profiles 
(for a recent example see \cite{poland}). Detailed knowledge of
competing RNA/RNA and RNA/DNA bindings is also becoming desirable for
cDNA microarrays \cite{deutsch}.

Previously we \cite{us} suggested that a deeper systematic framework lay 
beneath the semi-empirical rules of the NN approach, while in Ref.\cite{aft}
we proposed such a scheme and demonstrated its equivalence to another
\cite{xia} model of oligonucleotide RNA duplexes containing WC pairs.

The structure of the paper is as follows. After a brief outline of the
NN dimer method we refine our base-pairing model proposed in Ref. \cite{aft}.
The equivalence between the two approaches is then described in detail.
Finally two-dimensional models are fitted to empirical $\Delta G$ values for 
DNA and RNA/DNA hybrid duplexes and compared statistically to their dimer 
analogues.

\section{Nearest Neighbour model}
The Nearest-Neighbour (NN) model of duplex formation, e.g see Ref. 
\cite{xia} is based upon the assumption that once an initiation barrier, 
preventing the formation of a single  H-bonded pair, is overcome the 
duplex/single strand interface propagates along the strands ``zipping'' the 
two strands up into a duplex. The critical assumptions are:

1) The process occurs at fixed strand concentrations.

2) Derivation of thermodynamical parameters from melting curves is
for an assumed equilibrium between two (helix and coil) possible states.
 
3) The thermodynamic properties of the duplex depends linearly upon the 
frequencies of adjacent pairs of bases (dimers). In particular the formula 
used to estimate duplex free energy of formation
is
\begin{equation}
\Delta G = \Delta G_{init}+ \Delta G_{ssym} +\sum N_{WY/XZ} \Delta G_{WY/XZ}. \label{nn}
\end{equation}
Here $\Delta G_{init}$ is a free energy initiation step including
translational and rotational entropy loss. It is, in principle, both
(duplex) length- and sequence-dependent but for short duplexes is typically 
assumed to depend only upon the identity of the terminal base pairs.
$N_{WY/XZ}$ denotes the frequency with which dimer $5'WX3'/3'YZ5'$
occurs in the duplex while $\Delta G_{WY/XZ}$ is its free energy
contribution to the helix formation. The latter is interpreted as 
the dimer propagation energy associated with the zippering of strands.
Dominant contributions to this propagation energy are the van der Waals 
``stacking'' between adjacent base pairs and specificity-conferring H-bonding 
of complementary base pairs. 
Finally an extra entropy term is added for self-complementary
duplexes (i.e. helices formed from strands with identical sequences)
due to the extra twofold symmetry. Theoretically $\Delta G_{ssym}=RT \ln 2 \sim
0.43$ kcal mol$^{-1}$ at physiological temperature $T=310K$.

\subsection{Model parameters}
In the simplest instance one assumes that the dimer propagation energy is 
independent of ``zippering'' direction, that is, $\Delta G_{WY/XZ}=\Delta G_{ZX/YW}$ and therefore $N_{WY/XZ}=N_{ZX/YW}$. It is easy to check that if $p$ 
types of pair are possible then this symmetry reduces the number of types of
dimer from $(2p)^{2}$ to $p(2p+1)$. We shall refer to this below as the
``symmetric'' dimer approximation.

For oligomers, i.e., helices with formally defined ends, there are also $2p$ possible 
terminal parameters. In this case the hypothetical $5'-3'$ dimer symmetry must be 
compatible with the global $5'-3'$ symmetry of the full duplex, leading to $p$ 
constraints \cite{Gray1} upon the numbers of independent dimer ($N_{WX/YZ}$) and 
terminal ($N_{5'P/Q3'}$) frequencies which coexist in the model. For convenience $p$ of
 the terminal parameters are typically eliminated, e.g., $N_{5'P/Q3'}=N_{5'Q/P3'}$.
The maximal number of {\it independent} NN frequencies in the model is 
therefore
\begin{eqnarray*}
\nu = p+p(2p+1).
\end{eqnarray*}
Oligomer models derived for RNA \cite{xia} and DNA duplexes \cite{WC}, 
\cite{Gray2} containing Watson-Crick (WC) pairs ($p=2$) exhibit good agreement with 
experimental data for short, ``two state'' duplexes while accuracy is 
decreased for duplexes containing mismatches (Refs. \cite{GT}-\cite{self}
for DNA or see \cite{gu}-\cite{x2} for RNA. This decrease is generally attributed to 
the emergence of non-nearest neighbour effects such as pairing geometries with reduced 
complementarity.
The ``asymmetric dimer'' approximation, where the zipping direction is 
distinguished, has $(2p)^{2}$ dimer parameters and may be applied to
hybrid RNA/DNA duplexes or, possibly, as a more sophisticated model 
of ordinary (RNA/RNA or DNA/DNA) homoduplexes. In fact the asymmetric dimer 
model is required to obtain reasonable predictions for RNA/DNA hybrids 
(\cite{Gray2}, \cite{hyb1}, \cite{hyb2}) but provides only marginal 
improvements in predictions for 
homoduplexes \cite{xia}, \cite{aft}.

It is also known \cite{unif} for WC-paired DNA polymers that a model 
consisting of eight ``invariants'', or linear combinations of dimer steps,
provides predictions of comparable accuracy to the model including parameters for all ten dimers. Previously we suggested that the base-pairing 
model developed in Ref. \cite{aft} for RNA duplex formation is more 
naturally compared with the NN method at this independent, short 
sequence (ISS) level, which we elaborate upon below.

 \section{Base-pairing model}
Our model is motivated by the problem of including sequence-dependence in
dynamical, base-pair level descriptions of DNA \cite{review}.  It follows 
from the NN assumption (3) above;  if a physical property is a linear 
function of  the duplex NN dimer quantities, then there must be an equivalent 
expression which is quadratic in labels associated with individual bases. 

For the instance of thermodynamical quantities this quadratic description
corresponds to a systematic summation of  two-body correlations.
Given the pre-eminent role of van der Waals ``stacking'' and H-bonding 
interactions in helix stabilisation, these correlations might be 
representative of the interactions between, for example, amino/keto 
functional groups and heterocyclic rings.  
Moreover terms which are independent of sequence content variations, 
depending only upon overall sequence length have a natural interpretation as
contributions to the generic (B form, in the present case) helical backbone 
structure.

Consider the dimer propagation energy parameters
$\Delta G_{WY/XZ}$ associated with the ``zipping up'' of dimers 
$5'WY3'/3'XZ5'$. We shall assume that they may be decomposed quadratically as
\begin{eqnarray}
\Delta G_{WY/XZ} \equiv (W^{T},X^{T})\left( \begin{array}{cc}
P_{11} & P_{12} \\
P_{21} & P_{22} \end{array}\right)
\left( \begin{array}{c}
Y\\
Z\end{array}\right)  \label{corr}
\end{eqnarray}
Here the matrix entries $P_{ij}$ represent generic correlations
between the various sites of the dimer, while the vector entries 
$W$, $X$, $Y$, $Z$ are vectors encapsulating the sequence variation.
For lack of better nomenclature we shall refer to matrix elements as
``correlations'' and the vectors as ``weights''. Three observations
 are in order:

1) The diagonal entries $P_{ii}$ (we shall call them $h_{ii}$) are associated
with bases which are H-bonded while off-diagonal ones are associated with
``stacked'' neighbours (which we call $s=P_{12}$ and $t=P_{21}$).

2) $W$, $X$, $Y$, $Z$ are vectors in some abstract ``weight'' space. In order
to reproduce the correct number of matrix elements, the dimensionality, $p$
of this space must coincide \cite{aft} with the number of permissible base 
pairs, $d$ above. The nature of this weight space is discussed in a later section.

3) Given the H-bonding specificity between complementary pairs, the inter- and
intra-strand correlations are not independent of one another. The
stacking correlations are thus understood to contain contributions from both these types of interaction.

Therefore expression (\ref{corr}) can be rewritten in the form used previously \cite{aft}.
\begin{eqnarray*}
\Delta G_{x_{1}x_{2}/y_{1}y_{2}} =
(y^{T}_{1}, y^{T}_{2}) \left( \begin{array}{cc}
                             h_{1} & s \\
                             t & h_{2} \end{array} \right)
\left( \begin{array}{c}
x_{1} \\
x_{2} \end{array} \right),
 \label{NNform}
\end{eqnarray*}
where bases $x_{i}$ and $y_{i}$ refer to bases on the $5'-3'$ and $3'-5'$ 
oriented strands respectively and the Roman index $i$ refers to the location
of the base within the dimer. We shall also adopt the convention that Greek 
indices signify internal ``weight space'' degrees of freedom.

The distinction between symmetric and asymmetric dimer approximations is 
quantified by the existence of a $5'-3'$ symmetry transformation, $\Gamma_{2}$.
In the spatial point group of a dinucleotide dimer this transformation is
just a $C_{2}$ rotation about an axis perpendicular to the average plane of 
the (dimer) molecule passing through the centre of mass.
Given that this model is concerned with correlations between abstract labels
in a multi-dimensional vector space and not {\it a priori} in physical 3-D space  we can only require that the transformation be involutive, i.e., 
\begin{eqnarray*}
(\Gamma_{2})^{2}=1_{p}
\end{eqnarray*}
where $1_{p}$ denotes the $p\times p$ identity matrix. In order for expression
(\ref{NNform}) to be manifestly $5'-3'$ symmetric it must be the case that:
\begin{eqnarray}
h_{1}-(\Gamma_{2})^{T}h^{T}_{2}\Gamma_{2}&=&0, \label{53sh}\\
s-(\Gamma_{2})^{T}s^{T}\Gamma_{2}&=&0, \label{53ss}\\
t-(\Gamma_{2})^{T}t^{T}\Gamma_{2}&=&0. \label{53sy}
\end{eqnarray}
For any suitable (involutive) choice of $\Gamma_{2}$ it is readily found
that the number of independent $h$, $s$ and $t$ matrix elements is $p(2p+1)$ 
in agreement with the number of symmetric dimer parameters.
Without loss of generality, in the remainder of the paper  we shall therefore 
assume $\Gamma_{2}=-1_{p}$.

\subsection{Duplex model}
We now construct the correlations for a full duplex of $n$ base pairs 
$x_{1}\ldots x_{n}/y_{1} \ldots y_{n}$. As in the NN approach the duplex heat 
of formation is approximated by a summation of internal dimer propagation 
energies plus, for oligomers, terminal-environmental effects:
\begin{equation}
\Delta G =  \Delta G_{term/env} + 
\sum^{n-1}_{i=1}\Delta G_{x_{i}x_{i+1}/y_{i+1}y_{i}} \label{fd}
\end{equation}
Here, c.f. Eq (\ref{nn}), the summation is over individual sites within the 
duplex rather than dimer occurrence frequencies. Let us denote the duplex 
analogues of dimer submatrices $h$, $s$ and $t$ by capital letters. If the 
global $5'-3'$ transformation is written $\Gamma_{n}$ the global symmetry constraints are
\begin{eqnarray}
H_{i}-(\Gamma_{n})^{T}H^{T}_{\tilde{i}}\Gamma_{n}&=&0, \label{53dh}\\
S_{ij}-(\Gamma_{n})^{T}S_{\tilde{j}\tilde{i}}^{T}\Gamma_{n}&=&0 ;
\hspace{0.1cm} j>i, \label{53ds}\\
T_{ij}-(\Gamma_{n})^{T}T_{\tilde{j}\tilde{i}}^{T}\Gamma_{n}&=&0 ;
\hspace{0.1cm} j<i, \label{53dt}
\end{eqnarray}
with $i$ and $j$ running from $1$ to $n$ and where we have defined
the $5'-3'$ reflected index $\tilde{i}=n+1-i$.
Naturally in the NN approximation only those submatrices with indices $|i-j|\leq 1$ contain nonzero elements. Reconciling the global and dimer symmetries is 
therefore equivanent to substituting the symmetric dimer version of  (\ref{NNform}) in (\ref{fd}). In particular one finds \cite{aft}
\begin{eqnarray}
H_{1}=h_{1}=H^{T}_{n}, & & H\equiv H_{2}=h_{1}+h^{T}_{1}=H_{3}=\ldots =H_{n-1}, \label{reconc}\\
S_{i,i+1}= s, & & T_{i+1,i}= t. 
\end{eqnarray}
Note that for internal bases the H-bonding correlations effectively contribute
twice and in such a way that $H_{i}=H^{T}_{i}$ for $1<i<n$. This is desirable
due to the fact that constraint (\ref{53dh}) on its own would imply different
symmmetry properties for the central base pair(s) in odd- (even-) length 
duplexes. We note that this double counting of internal base pairs with 
respect to terminals may also be justified in terms of the relative 
propensities for fraying of the latter \cite{Ramreddy}.

Additional, environmental, perturbations to terminal bases may be formally 
incorporated by augmenting the original sequences with fictitious ``end neighbour'' vectors 
\cite{Goldstein}
\begin{eqnarray*}
(x_{1}, \ldots x_{n}) & \to  (e_{x},x_{1}, \ldots x_{n},e^{'}_{x}) \\
(y_{1}, \ldots y_{n}) & \to  (e_{y},y_{1}, \ldots y_{n},e^{'}_{y}).
\end{eqnarray*}
If these perturbations are themselves $5'-3'$ symmetric then the environmental 
effects may be incorporated as a linear term
\begin{equation}
\Delta G_{term/env}= \beta.(x_{1}+x_{n}+y_{1}+y_{n}),
\end{equation}
where $\beta$ is a constant, $p$-dimensional vector which is not required, 
i.e., vanishes in expressions for polymers or circular duplexes, if no
ends are formally defined.

\subsection{Number of Model Parameters}
Let us now recall the observation \cite{unif} that there is a more 
fundamental description of WC-paired polymers in terms of eight ``invariants''
and that moreover, when initiation terms are added, the predictions for
oligomers are comparable to those of the full model with ten dimer parameters.
 
Observe that bases $y_{i}$ and $x_{i}$ have, to this point, been effectively
treated as independent variables, but as is well known, the complementary 
H-bonding of nucleotide bases is highly specific. Thus we can assume that if 
the types of base-pairing occuring within the duplex are known then a 
relationship between complementary bases might be exploited. 
For example, if only WC pairing geometries occur in a given duplex then it 
follows that if site $y_{i}$ contains the base $C$, site $x_{i}$ must contain
a $G$. Thus the labels specifying one of the two bases are redundant.

In general for a base pair we may therefore write
\begin{equation}
y_{i}=\gamma_{i}x_{i}\equiv\lambda_{i} R_{i} x_{i}, \label{brot}
\end{equation}
where $\lambda>0$ is a scaling factor while $R$ is, in general, some 
length-preserving transformation and the matrices $\gamma_{i}$ are assumed 
to be characteristic of a particular pairing geometry. 
It follows from the expression
\begin{eqnarray}
y^{T}_{i} h x_{i}=  \lambda_{i} x_{i}^{T} R_{i}^{T}h x_{i}, \label{gam} 
\end{eqnarray}
that the off-diagonal entries of $(R^{T}h)$ only occur in linear combinations,
thereby reducing the number of independent dimer parameters by $p(p-1)/2$.

In the special case of  WC-pairing geometries ($p=2$) we therefore have 9 such 
parameters. To obtain a model with eight parameters we require a 
further assumption about the form of $R_{i}$. From (\ref{reconc}) the 
internal H-bonding correlations are symmetric and it is easy to see that the
four elements $H^{\alpha\beta}$ occur in just two linear 
combinations precisely when $R$ has zero values along its' diagonal.
This effective eight-parameter treatment of internal base pairs is directly
comparable to the description of polymers, i.e. sequences with no mormal ends,
in terms of eight ``invariants'' or ISS.

Now consider the difference between the contributions of $5'P/3'Q$ and 
$5'Q/3'P$ pairs. If the pairs are internal, since $H=H^{T}$ the H-bonding 
enthalpies are equal: 
\begin{eqnarray*}
q^{T}Hp=p^{T}Hq.
\end{eqnarray*}
It follows that any orientation effects are manifest only at the 
terminals where
\begin{eqnarray*}
q^{T}hp=p^{T}h^{T}q\neq p^{T}hq
\end{eqnarray*}
Thus in the WC case the difference $h_{12}-h_{21}$ is a measure of the 
magnitude of these effects. If, as is commonly assumed in the NN approach for 
DNA and RNA duplexes, $5'G/3'C$ and $5'C/3'G$ terminal pairs are equivalent 
(and similarly for A.T/U) this quantity should be a small correction. 
Therefore the model with eight dimer parameters obtained by assuming $h_{12}\simeq h_{21}$ should be a reasonable approximation of the full, 9 parameter 
version, which we verify in the results below.
Analogously in Ref. \cite{unif} the ability of the ``rank deficient'' DNA 
polymer model to reproduce DNA oligomer data was rationalised thus:
``...most of the sequence dependence of oligonucleotide DNA thermodynamics is 
captured in the first eight terms and the remaining two are small perturbations...'' 

In two dimensions there are just two candidates for the transformations $R_{i}$ defined in (\ref{gam}), these are proportional to 
\begin{eqnarray*}
\sigma_{1}=\left( \begin{array}{cc}
                 0 & 1 \\
                -1 & 0 \end{array} \right), & & 
\sigma_{2}=\left( \begin{array}{cc}
                 0 & 1 \\
                 1 & 0 \end{array} \right).
\end{eqnarray*}
If, for simplicity the scalings $\lambda_{i}=1$, the latter form can be 
rejected since  $\sigma_{2}=\sigma_{2}^{-1}$.
This would ensure that the six stackings $s^{\alpha\beta}$, $t^{(2-\beta)(2-\alpha)}$ always appear in just three linear combinations, reducing the 
number of independent parameters from 8 to 5.
For $\gamma=\pm \sigma_{1}$ the property $\gamma^{-1}=-\gamma$ means that 
the relative signs between $s$ and $t$ elements are sequence-dependent, 
thereby ensuring the number of model parameters is fixed at eight.

\subsection{Equivalence with dimer model}
For clarity we now show the equivalence of the present model 
with the NN approach for canonical duplexes. Firstly the duplex pairing 
matrix is written in terms of Hydrogen-bonding and stacking components:
\begin{eqnarray}
\Delta G(X,Y)& =& \Delta G_{HB}+\Delta G_{ST}, \label{scnn}\\
\Delta G_{HB}&=& y^{T}_{1}hx_{1} + y^{T}_{n}hx_{n} + \sum_{i=2}^{n-1}
y^{T}_{i}H x_{i}, \label{hbnn} \\
\Delta G_{ST} &= &\sum_{i=1}^{n-1}\left( 
y^{T}_{i} s x_{i+1} + y^{T}_{i+1} t x_{i} \right). \label{stnn}
\end{eqnarray}
Now consider a dimer $5'XZ3'/3'WY5'$ occuring with frequency $N_{XZ}$ in a 
given duplex. For symmetric dimers $N_{XZ}=N_{YW}$ and the stacking term 
$\Delta G_{ST}$ depends on ten linear cominations of these 16 frequencies:
\begin{eqnarray}
\Delta G_{ST} & \equiv & 2N_{GC} S_{GC/GC}+2N_{CG} S_{CG/CG}+2N_{AT} S_{AT/AT}+2 N_{TA} S_{TA/TA} \nonumber \\
& & +(N_{GA}+N_{TC}) S_{GA/TC} + (N_{GT}+N_{AC}) S_{GT/AC}+(N_{CA}+N_{TG}) S_{CA/TG}\nonumber \\
& & +(N_{AG}+N_{CT}) S_{AG/CT}+(N_{AA}+N_{TT}) S_{AA/TT}+(N_{GG}+N_{CC}) S_{GG/CC} \label{gst} \\
S_{XZ/WY}&=& \sum_{\alpha,\beta=1}^{p}(s^{\alpha\beta} x_{\alpha}y_{\beta} +
t^{\alpha\beta} w_{\alpha}z_{\beta}). \nonumber
\end{eqnarray}
Similarly, if $h=H/2$, the H-bonding terms contain four numbers:
\begin{equation}
\Delta G_{HB}  =  (n^{t}_{GC}+ 2n^{i}_{GC}) h_{GC} + (n^{t}_{AT}+2n^{i}_{AT})
h_{AT}; \hspace{0.2cm} H_{xy}\equiv \sum_{\alpha,\beta=1}^{p} h^{\alpha\beta}y_{\alpha}x_{\beta}, \label{ghb}
\end{equation}
where $n^{t}_{GC}$, $n^{i}_{GC}$, $n^{t}_{AT}$, and $n^{i}_{AT}$ denote the 
numbers of terminal and internal G.C and A.U pairs respectively.
These numbers of base pairs are not independent of the stacking frequencies
$N_{XY}$ however:
\begin{eqnarray*}
(n^{t}_{GC}+ 2n^{i}_{GC})&=&N_{GA}+N_{GT}+2N_{GG}+2N_{CC}+N_{AG}+N_{TG}+2N_{CG}, \\
(n^{t}_{AT}+ 2n^{i}_{AT})&=&2N_{AA}+2N_{AT}+N_{AG}+N_{AC}+N_{GA}+2N_{TA}+N_{CA}.
\end{eqnarray*}
Combining these identities in (\ref{ghb}) with (\ref{gst}) one sees that the
coefficients of $N_{XY}$ in (\ref{scnn}) contain both H-bonding and stacking 
correlations. In this way one obtains ten dimer parameters equivalent to 
those of the Interacting Nearest Neighbour with H-Bonding (INNHB) \cite{xia}
(see Table \ref{dims} in the results for verification of this).
Furthermore the ten dimer parameters in our model are just linear 
combinations of the eight matrix elements, similar to the eight ISS parameters
discussed in Ref. \cite{unif}.

\subsection{Weight space}
Having discussed the 2-body correlation matrix in detail we now turn
to the vector ``weight'' space. In order to obtain a model which is 
equivalent to the polymer dimer model we have imposed just two ``rules'' on 
the weight space:

(1) The vectors for complementary bases are orthogonal and of equal length.

(2) The model of duplexes with WC pairing must have $p=2$, therefore
the vectors for G,C,A,T all live in the same (two-dimensional) space.

Note that there are still $2p$ unknown parameters (the coordinate values 
for, say G and A) however, rather than attempt to obtain them from a
fit to empirical data assumptions may be made about what the vector
coordinates represent. 

For WC pairs in ref. \cite{aft} we assumed one coordinate counted 
the number of H-bonds formed, the other indicated whether the the heterocyclic
ring was purine or pyrimidine. With the basis
\begin{eqnarray}
\{ G,C,A,U(T) \}=\{(\sqrt{3},1/2),(-1/2,\sqrt{3}),(-\sqrt{2},1/2),(-1/2,-\sqrt{2}\}, \label{basis}
\end{eqnarray}
we obtained an 8-parameter, so-called ``rank-deficient'', model of RNA 
oligonucleotides statistically identical to the conventional dimer model 
(e.g. Refs. \cite{xia}, \cite{WC}). To obtain fitted models from DNA/DNA and 
RNA/DNA data below we shall assume that the same degrees of freedom (\ref{basis}) contribute to WC pairing. The effects the of differing sugar backbone
geometries will be manifested in the relative magnitudes of the fitted coupling
values for the various duplexes.

\section{Results: WC Pairs in DNA}
The next stage of our comparitive analysis is to obtain the DNA pairing model 
from data for duplexes with two-state melting transitions and consisting of
only Watson Crick pairs.
Following Refs \cite{WC}, \cite{Gray2} we construct models for thermodynamic 
parameters $\Delta G$, $\Delta H$, $\Delta S$ for two sets of sequences
one set 44 of duplexes terminated by GC pairs only, the other containing,
in addition to these 44, eight duplexes with at least one AT terminal.

We use an initiation term with separate $5'T/3'A$ and $3'T/5'A$ parameters, 
consistent with the observation of SantaLucia {\it et al.} \cite{WC} that the
former have a tendency to fray:
\begin{equation}
\Delta G_{init}=\alpha_{1} n^{t}_{GC} +
+\alpha_{2} n^{t}_{5'T}+
\alpha_{3} n^{t}_{5'A}. \label{initnn}
\end{equation}
The self-complementary entropy ``penalty'' $\Delta G_{sym}$ is set to the t
heoretical value $0.43$ kcal mol$^{-1}$ at a temperature of 310K. The form of 
our fitting function is thus given by (\ref{initnn}) plus (\ref{hbnn}) and 
(\ref{stnn}):
\begin{eqnarray}
\Delta G(X,Y)& = & \Delta G_{init}+\Delta G_{sym} +
y^{T}_{1} h x_{1}+ y^{T}_{n} h^{T}x_{n} \nonumber \\
& & \mbox{}+ \sum_{i=2}^{n-1} y^{T}_{i}(h+h^{T})x_{i}
+\sum_{i=1}^{n-1} (y^{T}_{i}sx_{i+1} + y^{T}_{i+1}tx_{i}). \label{fft1} 
\end{eqnarray}
Here $h$, $s$ and $t$ are $2\times 2$ matrices while vectors $x_{i}$, $y_{i}$
take the appropriate values from the base vector set (\ref{basis}).
Due to the assumed $5'-3'$ dimer symmetry $s$ and $t$ are symmetric, while
we have kept the distinction between $h$ and $h^{T}$ for the purpose of 
comparing rank-deficient and rank-determinant parameter sets.

To compare the NN dimer and base-pairing approaches we shall compute the
same statistical parameters used in other studies, for example, Ref.\cite{xia}.
In addition to the root of the mean of squared residuals $\sigma$ we include
the unweighted $\chi^{2}$ parameter is computed to be
\begin{equation}
\chi^{2}=\sum \left(\frac{G_{p}-G_{o}}{\sigma}\right)^{2}, \label{chsq}
\end{equation}
where $G_{p}$, $G_{o}$ and $\sigma$ are the predicted and observed values of 
G and the rms value respectively. Here $f$ is the number of 
observations less the number of model parameters.
The reduced parameter $\chi^{2}/f$ should have value close to unity for
a good fit. The $Q$ value estimates the likelihood of obtaining a particular 
value of $\chi^{2}$ by chance:
\begin{equation}
	Q=\Gamma (f/2,\chi^{2}/2)/ \Gamma (f/2)
	\label{qval}
\end{equation}
where $\Gamma (a)$ and $\Gamma(a,z)$ denote the complete and incomplete gamma
functions respectively. Small $Q$ values signify that discrepancies between 
the model predictions and experimental data are unlikely to be due to chance.

Using Eqs (\ref{chsq}), (\ref{qval}) we compute these statistics for the
predicted values of $\Delta G$ in six models in Table \ref{wcstats}. 
The models NN1 and NN2 are respectively
the 11- and 12-parameter models obtained in Ref. \cite{WC} for sets of
G.C- and G.C/A.T- terminated duplexes. JB1a and JB1b denote, respectively
the data for rank-determinant ($h=h^{T}$) and rank deficient $(h\neq h^{T})$
models for the 44 GC terminated duplexes. Similarly JB2a and JB2b denote
the same models for the full set of 52 sequences. 
\begin{table}[htb]
\begin{center}
    \begin{tabular}{cccc}
\hline \\
 & $\sigma^{a}$ & $\chi^{2}/f$ & $Q$ \\
\hline
NN1 & 0.33 & 1.22 & 0.10 \\
JB1a & 0.33 & 1.26 & 0.14 \\
JB1b & 0.33 & 1.22 & 0.17 \\
NN2 & 0.35 & 1.30 & 0.10 \\ 
JB2a & 0.34 & 1.28 & 0.12 \\
JB2b & 0.34 & 1.24 & 0.14 \\
\hline \\
\end{tabular} \label{wcstats}
\caption{Comparison of statistics for the standard Nearest-Neighbour (NN)
parameters and the fitting of our models to the same data. $^{a}$Units are
kcal mol$^{-1}$. }
\end{center}
\end{table}
Several observations may be made immediately from Table \ref{wcstats}.
Firstly, the fit rms values for all models are in good agreement, however
if the distinction between AT and TA terminal parameters is removed the rms
values of JB2a, JB2b would rise respectively to 0.41 and 0.39 kcal mol$^{-1}$.
Note however that introducing separate parameters for $5'G/3'C$ and $5'C/3'G$
in all instances provides slight improvements in $\sigma$ values
of $\leq 0.005$ kcal mol$^{-1}$ while increasing $\chi^{2}/f$ and decreasing
$Q$. Therefore the optimal choice of initiation term (\ref{initnn}) is 
validated.

The parameter set for estimating $\Delta G$ for the best model, JB2a 
is given by (units are kcal mol$^{-1}$)
\begin{eqnarray}
 \alpha_{1}= 0.833, & \alpha_{2}= 0.98 & \alpha_{3}= 1.84 \nonumber\\
h=\left(\begin{array}{cc}
0.139 & -0.293 \\
-0.305 & -0.139 \end{array}\right), & 
t=\left(\begin{array}{cc}
-0.071 & -0.033 \\
-0.033 & 0.180 \end{array}\right), & 
s=\left(\begin{array}{cc}
-0.190 & 0.019 \\
0.019 & 0.055 \end{array}\right). \label{wcf2}
\end{eqnarray}
Of course the models should also be compared at the level of dimer 
propagation energies. These quantities are found from (\ref{fft1}) via
\begin{eqnarray*}
\Delta G_{XZ/YW} \equiv X^{T}hY + Z^{T}hW + Y^{T}sZ+W^{T}tX
\end{eqnarray*}
and using the basis (\ref{basis}) and parameters (\ref{wcf2}). 
Values are compared to the dimer quantities of Ref.\cite{WC} in Table \ref{dims} below.
\begin{table}[htb]
\begin{center}
    \begin{tabular}{cccc}
\hline \\
$\Delta G$ & NN$^{a}$  & BP$^{b}$ \\
\hline
GG/CC & $-1.77\pm 0.06$ & -1.76 \\
GC/CG & $-2.28\pm 0.08$ & -2.28 \\
CG/GC & $-2.09\pm 0.07$ & -2.09 \\
AA/TT & $-1.02\pm 0.04$ & -1.01 \\
AT/TA & $-0.73\pm 0.05$ & -0.77 \\
TA/AT & $-0.60\pm 0.05$ & -0.63 \\
GA/CT & $-1.46\pm 0.0$5 & -1.52 \\
GT/CA & $-1.43\pm 0.05$ & -1.37 \\
AG/TC & $-1.16\pm 0.07$ & -1.13 \\
TG/AC & $-1.38\pm 0.06$ & -1.36 \\
\hline\\
\end{tabular} \label{dims}
\caption{Comparison of NN symmetric dimer parameters with those derived from 
the rank-deficient parameters (\ref{wcf2}). All units are kcal mol$^{-1}$.
$^{a}$Dimer parameters from 12-parameter model obtained by SantaLucia 
{\it et al.} \cite{WC}. $^{b}$BP denotes the base pairing model with rank deficient
parameter set JB2b.}
\end{center}
\end{table}
For completeness the fitted matrix elements for $\Delta H$ and $\Delta S$
and related dimer quantities are included in the appendix.

\section{Results: WC pairs in RNA/DNA hybrids}
We now modify the approach in order to analyse thermodynamic data for
RNA/DNA hybrids, in particular the 68 sequences used by Sugimoto and co-workers\cite{hyb2}, and compare our results 
to previous NN analyses \cite{hyb2}, \cite{Gray2}.

The major difference to DNA or RNA hybrids is that, owing to the different
backbones of the strands the zippering direction is readiliy distinguished;
there are no local dimer or global duplex symmetries, nor a self-symmetric 
entropy term. Indeed a naive fit of 
the symmetric dimer fitting function \ref{fft1} yields an rms value of 0.57 
kcal mol$^{-1}$, considerably poorer to the homoduplex models. 

The dimer propagation matrix is therefor given by (\ref{NNform}) where
constraints (\ref{53sh}-\ref{53sy}) do not hold and the number of independent
matrix elements is simply $(2p)^{2}$. The comparison of dimer and global 
correlations now yields
\begin{eqnarray}
H_{1}=h_{1}, & & H_{n}= h_{2}, \nonumber\\
H\equiv H_{2}=h_{1}+h_{2}=H_{3}=&\ldots& =H_{n-1}, \label{rc2}\\
S_{i,i+1}= s, & & T_{i+1,i}= t . \nonumber
\end{eqnarray}
Hence the fitting function is now given by
\begin{eqnarray}
\Delta G(X,Y)& = & \Delta G_{init} +
y^{T}_{1} h_{1} x_{1}+ y^{T}_{n} h_{2}x_{n} \nonumber \\
& & \mbox{}+ \sum_{i=2}^{n-1} y^{T}_{i}(h_{1}+h_{2})x_{i}
+\sum_{i=1}^{n-1} (y^{T}_{i}sx_{i+1} + y^{T}_{i+1}tx_{i}). \label{fft2} 
\end{eqnarray}
where, in contrast to the single initiation term of Ref \cite{hyb2} we shall 
assume distinct terminal parameters for the terminii
\begin{eqnarray*}
r(G)/d(C)\equiv r(C)/d(G) & & r(A)/d(T)\equiv r(U)/d(A). 
\end{eqnarray*}
Of course in general four such parameters may be distinguished, however 
initially we shall consider an initiation energy form 
\begin{eqnarray*}
\Delta G_{init} =\alpha_{1} n^{GC}_{t} + \alpha_{2} n^{AT(U)}_{t}.
\end{eqnarray*}
The resulting model has, naively $2+16$ parameters to be compared
with $1+16$ for the NN model \cite{hyb2}. However RNA/DNA hybrids, like
homoduplexes exhibit high complementarity in H-bonding.
Since the ``weight vectors'' (\ref{basis}) have been shown to capture the
essential sequence-dependent interactions of both DNA/DNA and RNA/RNA 
\cite{aft} helix formation and, noting that the RNA/DNA hybrid also has a 
regular helical geometry, it is reasonable to investigate whether they are 
successful in the latter instance.

With the existence of complementarity transformations (\ref{gam}), two
of each of the $h_{1}$ and $h_{2}$ matrix entries appear in single linear 
combinations. 
For the DNA/DNA model we obtained a rank-deficient model by neglecting
the distinction between the H-bond correlations of the two terminii. The
analogy in the hybrid case is to suppose $h_{1}\simeq h_{2}$, the validity
of which we check below. The rank-deficient parameter set for hybrids
therefore has 2+11 parameters, an improvement of 4 over the NN model.

In fact we find that this model reproduces the empirical $\Delta G$ data 
(rms 0.38 kcal mol$^{-1}$) slightly better than the NN models \cite{hyb2}, 
\cite{Gray2} and in addition gives better reduced $\chi^{2}/f$ and $Q$ values, 
see Table \ref{wcstats2}. The fitted values obtained (in kcal mol$^{-1}$) are
\begin{eqnarray}
 \alpha_{1}= 1.65, & \alpha_{2}= 1.52 & \nonumber\\
h=\left(\begin{array}{cc}
-0.151 & 0.277 \\
0.369 & 0.151 \end{array}\right), & 
t=\left(\begin{array}{cc}
-0.035 & -0.065 \\
 0.002 & -0.011 \end{array}\right), & 
s=\left(\begin{array}{cc}
0.010 & 0.060 \\
0.078 & 0.035 \end{array}\right). \label{wc16a}
\end{eqnarray}
Sugomoto and coworkers used a single helix initiation parameter of 3.1 kcal 
mol$^{-1}$ which is consistent with our initiation term which takes the
values $3.05$, $3.17$ or $3.30$ kcal mol$^{-1}$ depending on whether the 
duplex has respectively 0, 1 or 2 $G.C$ terminals.

Several other observations may be made about the matrix elements (\ref{wc16a}).
Firstly the difference $h_{12}-h_{21}$ is roughly four times larger in the 
hybrid case, therefore the model already incorporates significant distinctions
between $5'A/3'T$ and $5'T/3'A$ (and similarly for GC) terminii.
Doubling the number of initiation terms therefore leads to only a small
increase in accuracy for the cost of two extra parameters.

More surprising is the result that $s_{11}\simeq -t_{22}$ and 
$s_{22}\simeq -t_{11}$ would provide excellent approximations, 
reducing the parameter number further to 11. In the absence of an obvious
theoretical reason for this coincidence of stackings we cannot reject 
the possibility that it is an artifact of the data. The rank deficient 
parameter set, denoted H1 in Table \ref{wcstats2}, for RNA/DNA hybrids 
therefore contains 13 parameters.
Finally we consider the 16 parameter rank-determinant ($h_{1}\neq h_{2}$) 
model, denoted H2 in Table \ref{wcstats2}. As anticipated it is found to 
provide a marginal improvement in $\sigma$ (of $\leq 0.005$ kcal mol$^{-1}$)
but at a cost reflected in the reduced $\chi^{2}$ and $Q$ statistics.
Again for completeness, in the appendix we tabulate fitted parameters for
$\Delta H$ and $\Delta S$ parameters.
\begin{table}[htb]
\begin{center}
    \begin{tabular}{cccc}
\hline \\
$\Delta G$ & $\sigma^{a}$ & $\chi^{2}/f$ & $Q$ \\
\hline
NN$^{b}$ & 0.45 & 1.57 & 0.01 \\
H1 & 0.38 & 1.24 & 0.12 \\ 
H2 & 0.38 & 1.31 & 0.07 \\
\hline \\
\end{tabular} \label{wcstats2}
\caption{Comparison of statistics for models of hybrid duplex formation.
$^{a}$Units are kcal mol$^{-1}$. $^{b}$NN data was calculated using model 
parameters (NN1) of Sugimoto {\it et al.} \cite{hyb2}. 
H1 and H2 denote rank deficient and rank determinant base-pairing parameter
sets respectively.}
\end{center}
\end{table}

\section{Conclusion}
In this paper we have refined the idea, first presented in Ref. \cite{aft}, 
that a description of duplex formation in terms of (fictitious) two-body 
correlations is equivalent to the more commonly known NN dimer method.

For all three cases (RNA, DNA and RNA/DNA hybrid helices) fits to 
empirical data confirm that this is indeed the case.
Moreover the connection between rank-deficient polymer and rank-determinant
oligomer NN parameter sets, known for DNA (see, e.g., Ref. \cite{unif}) is
transparent via our approach and enables similar approximations for RNA 
\cite{aft} and RNA/DNA hybrids to be made.

In the instances of RNA and DNA homoduplexes while the former approximation 
does not offer an increase in model accuracy, statistics of model significance
(e.g. $\chi^{2}$, $Q$) suggest the base-pairing approach to be more 
favourable on the grounds that it uses fewer and more ``fundamental'' model 
degrees of freedom. For RNA/DNA hybrids an improvement in both accuracy
and parameter numbers is also observed over the NN method, indeed the high
$\chi^2$ and low $Q$ probability for the latter in Table \ref{wcstats2} 
strongly suggest a good deal of redundancy in using the dimer energy 
parameters.

The base-pairing model also suggests the interesting possibility of a
kind of universality forstabilising interactions in Watson-Crick paired
helices. Specifically the same numerical weights (\ref{basis}) are found to 
encapsulate the sequence-dependence for all three types of helices, the 
differences between fitted correlation matrix elements being attributed to 
the effects of different helical geometries (A-type for RNA/RNA and RNA/DNA, 
B for DNA/DNA). 

It should be emphasised that the base-pairing model is only a description
of the sequence-dependence of helical oligonucleotides in terms of
two-body correlations labelled by ``weight vectors''.
However the suggestion of a connection between this picture with an effective 
interaction potential in three spatial dimensions is appealing. The nature of 
these ``weight vectors'' is not yet clear, and further insight may be gained 
from extending the method to mismatch pairing geometries.

\vspace{1cm}
\noindent
\large{ {\bf Acknowledgements}}

\noindent
This research was funded by Australian Research Council grant DP0344996.
JB wishes to thank T Ramreddy for a helpful discussion.

\section{Appendix}
\begin{eqnarray*}
h_{RNA}=\left(\begin{array}{cc}
0.122 & -0.232 \\
-0.232 & -0.122 \end{array}\right), & 
t_{RNA}=\left(\begin{array}{cc}
0.156 & -0.051 \\
-0.051 & -0.022 \end{array}\right), & 
s_{RNA}=\left(\begin{array}{cc}
0.070 & 0.067 \\
0.067 & -0.159 \end{array}\right),
\end{eqnarray*}

\end{document}